\documentclass[9pt,twocolumn]{revtex4-1}

\usepackage{graphicx}
\usepackage{float}
\usepackage{hyperref}
\usepackage{amsmath}
\graphicspath{ {../images/} }

\begin{document}
\title{Passive broadband full Stokes polarimeter using a Fresnel cone}

\author{R. D. Hawley}
\author{J. Cork}
\author{N. Radwell}
\author{S. Franke-Arnold}

\affiliation{SUPA and School of Physics and Astronomy, University of Glasgow, Kelvin Building, Renfrewshire, Glasgow, G12 8QQ, UK}
\affiliation{Corresponding author: r.hawley.1@research.gla.ac.uk}

% To be edited by editor
% \doi{\url{http://dx.doi.org/10.1364/optica.XX.XXXXXX}}

\begin{abstract}
Light’s polarisation contains information about its source and interactions, from distant stars to biological samples. Polarimeters can recover this information, but reliance on birefringent or rotating optical elements limits their wavelength range and stability. Here we present a static, single-shot polarimeter based on a Fresnel cone - the direct spatial analogue to the popular rotating quarter-wave plate approach. We measure the average angular accuracy to be $2.9^\circ$ ($3.6^\circ$) for elliptical (linear) polarisation states across the visible spectrum, with the degree of polarisation determined to within $0.12$ ($0.08$). Our broadband full Stokes polarimeter is robust, cost-effective, and could find applications in hyper-spectral polarimetry and scanning microscopy.

\end{abstract}

\maketitle

\section{Introduction}
Polarimeters identify the polarisation state of light and are an important tool in astronomy~\cite{Costa2001,Lites2008a,Sterzik2012a}, material characterisation~\cite{Johnson2000a,Kovaleva2004a,Losurdo2009}, remote sensing~\cite{Cloude1996a,Diner2007a,Imazawa2016} and medicine~\cite{Greenfield2000a,Ghosh2011b,Vizet2016}. Polarisation states describe two orthogonal complex field components and are commonly expressed as four real numbers in a Stokes vector~\cite{Bickel1985b}. Characterisation therefore requires at least four measurements, typically performed by either spatial splitting or temporal modulation. 

Spatial splitting techniques divide incident light into several sub-beams, each requiring its own polarisation analysing optics. More recently this has been implemented using "division of focal plane" spatial modulation, where the detector is divided with components made using advanced fabrication techniques~\cite{Nordin1999,Myhre2012a,Deng2017a}. Spatial splitting allows measurements to be obtained simultaneously, however often at the cost of complex or expensive setups. Recent devices address some of these issues~\cite{Chang2014d,BalthasarMueller2016,Estevez2016}. 

Temporal modulation techniques require taking sequential measurements, such as the rotating wave plate technique~\cite{Berry1977a} typically found in commercial polarimeters, liquid crystal variable retarders~\cite{Bueno2000b} or photo-elastic modulators~\cite{Liu2006b,Arteaga2012a}. These techniques become ineffective if the initial polarisation is varying on time-scales comparable to the modulation time. Moreover, devices with moving parts are prone to instability and broadband operation remains a challenge.

In this paper, we demonstrate a polarimeter based on the back-reflection from a Fresnel cone which is a direct spatial analogue to the popular rotating quarter-wave plate device~\cite{Berry1977a,Pelizzari2001a, Lin2012a}. We show that our polarimeter is intrinsically broadband and stable due to its lack of both birefringent elements and moving parts.

%Furthermore, optimisation techniques previously developed for the rotating quarter-wave plate polarimeter~\cite{Sabatke2000b,Dlugunovich2001a,Tyo2002a,Flueraru2008a,Romerein2011b} can be directly implemented in our system.

Fresnel cones - solid glass cones with a $90^\circ$ apex angle - act like azimuthally varying wave plates~\cite{Radwell2016}. Fresnel's equations predict that total internal reflection (TIR) produces phase shifts between the s and p polarisation components which are dependent on the angle of incidence and the refractive index of the cone material. The conical surface leads to an azimuthally varying decomposition of the input polarisation state into s and p components, and the resulting azimuthally varying phase shifts produce polarisation structures. Fresnel cones are simple to implement, intrinsically broadband and cost-effective.

Every spatially uniform initial polarisation state is mapped onto a unique polarisation structure by the Fresnel cone. Detection of this polarisation structure therefore allows us to identify the initial polarisation state. In the following we present the theoretical treatment of the Fresnel cone polarimeter as the spatial analogue of the rotating quarter-wave plate polarimeter and report on the performance and calibration of the device, before discussing further optimisation.

\section{Polarimeter Theory}
\label{PolTheory}

One of the most wide-spread and simple polarimeters uses a rotating quarter-wave plate and a linear polariser followed by an intensity measurement on a photodiode, which we summarise in the following. Written in the familiar Stokes formalism this is
\begin{equation}
\textbf{S'}=\textbf{M}_{\rm pol}\textbf{R}(-\theta)\textbf{M}_{\rm qwp}\textbf{R}(\theta)\textbf{S},
\label{QWP}
\end{equation}
where $\textbf{S}=[s_0,s_1,s_2,s_3]^T$ and $\textbf{S'}=[s'_0,s'_1,s'_2,s'_3]^T$ are the initial and final Stokes vectors respectively, the $\textbf{M}$ matrices are Mueller matrices corresponding to a horizontally aligned linear polariser ($\textbf{M}_{\rm pol}$), a quarter-wave plate with horizontal fast-axis ($\textbf{M}_{\rm qwp}$), and $\textbf{R}(\theta)$ is the Mueller rotation matrix by an angle $\theta$. The terms $\textbf{R}(-\theta)\textbf{M}_{\rm qwp}\textbf{R}(\theta)$  represent a quarter-wave plate rotated by an angle $\theta$ from horizontal (see Supplement 1). Here we are using a right-handed Cartesian coordinate system, where the light propagates in the $+z$ direction and the observer is looking towards the light source. The intensity $I(\theta)$ measured on the photodiode is then determined by the first row of~\autoref{QWP} as 
\begin{multline}
I(\theta)={s}_{0}'=\frac{1}{2} \bigg[ s_{0}+\frac{1}{2} s_{1}(1+\cos{(4\theta)})-s_{3}(\sin{(2\theta)}) \\
+\frac{1}{2} s_{2}(\sin{(4\theta)}) \bigg].
\label{RotQWP}
\end{multline}
Measurements for different $\theta$ values are taken by spinning the quarter-wave plate and recording the intensity as a function of time ($I(t)$), which is mapped to $I(\theta)$ through knowledge of the rotational frequency. \autoref{RotQWP} can be converted to a truncated Fourier series in $\theta$ and allows us to express the Fourier coefficients in terms of the components of the initial Stokes vector $\textbf{S}$. Measurement of these coefficients then allows recovery of $\textbf{S}$ through matrix inversion~\cite{Dlugunovich2001a,Flueraru2008a,Romerein2011b}. 

Our polarimeter approach is analogous to the above rotating quarter-wave plate technique. Our system consists of a beam-splitter, Fresnel cone, linear polariser and camera (shown in \autoref{fig:setup}), the equation for this system is
\begin{equation}
\textbf{S'}=\textbf{M}_{\rm pol}\textbf{R}(-\theta)\textbf{M}_{\rm wedge}\textbf{R}(\theta)\textbf{S},
\label{Cone}
\end{equation}
where $\textbf{R}(-\theta)\textbf{M}_{\rm wedge}\textbf{R}(\theta)$ are the Mueller matrices representing the Fresnel cone, comprised of double total internal reflection from a rotated ($\textbf{R}(\theta)$) glass wedge ($\textbf{M}_{\rm wedge}$). It is obtained by conversion of the Jones matrix found in~\cite{Radwell2016} following the method described in~\cite{Azzam1987}. For an appropriately chosen refractive index of the cone, formally $\textbf{M}_{\rm wedge}=\textbf{M}_{\rm qwp}$ and \autoref{Cone} becomes \autoref{QWP} where $\theta$ is now the azimuthal spatial angle with the origin of the polar coordinate system at the cone apex. The cone system is therefore revealed to be a direct spatial analogue to the rotating quarter-wave plate system.

\section{Polarisation state recovery}

For a polarimeter with optimally designed, ideal optical components, we can recover the initial Stokes vector \textbf{S} by solving \autoref{RotQWP} evaluated for at least four $\theta_i$. For this we re-express \autoref{RotQWP} as a truncated Fourier series in $\theta_i$:
\begin{equation}
I(\theta_i)=\frac{1}{2} \bigg[a_0+b_2\sin2\theta_i+a_4\cos4\theta_i+b_4\sin4\theta_i \bigg],
\label{startAnalysis}
\end{equation}
where the discrete Fourier coefficients for $N$ discrete angles are
\begin{align}
a_0=s_0+\frac{s_1}{2}&=\frac{2}{N}\sum_{i=1}^{N}I_i\\
b_2=-s_3&=\frac{4}{N}\sum_{i=1}^{N}I_i\sin 2\theta_i\\
a_4=\frac{s_1}{2}&=\frac{4}{N}\sum_{i=1}^{N}I_i\cos 4\theta_i\\
b_4=\frac{s_2}{2}&=\frac{4}{N}\sum_{i=1}^{N}I_i\sin 4\theta_i.
\end{align}
The values on the right hand side can be experimentally determined from Fourier analysis of the signal $I(\theta)$. The matrix form of Equations 5-8 is
\begin{align}
\underbrace{
\begin{bmatrix}
 1&\frac{1}{2}  &0  &0 \\ 
 0&  0&  0& -\frac{1}{2}\\ 
 0&  \frac{1}{2}& 0 &0 \\ 
 0&0  &\frac{1}{2}  &0 
\end{bmatrix}
}_{\textbf{M}}
\underbrace{
\begin{bmatrix}
s_0\\ 
s_1\\ 
s_2\\ 
s_3
\end{bmatrix}
}_{\textbf{S}}
=
\underbrace{
\begin{bmatrix}
a_0\\b_2\\a_4\\b_4
\end{bmatrix}
}_{\textbf{C}}
\end{align}
and the polarisation state \textbf{S} is then found by matrix inversion via
\begin{equation}
\textbf{S}=\textbf{M}^{\textbf{-1}} \textbf{C}.\label{endAnalysis}
\end{equation}
This analysis holds equally for an ideal quarter-wave plate polarimeter and our Fresnel cone polarimeter. In both cases optical elements may introduce unwanted polarisation shifts, however these can be accounted for with suitable analysis, producing optimal polarimeter results even for imperfect devices.

We find that the main source of unwanted polarisation shifts in the Fresnel cone polarimeter arises from the beam-splitter. To counteract these we include the Mueller matrices of the beam-splitter transmission ($\textbf{B}_{\rm trans}$) and reflection ($\textbf{B}_{\rm refl}$):
\begin{equation}
\textbf{S'}=\textbf{M}_{\rm pol} \textbf{B}_{\rm refl} \textbf{R}(-\theta) \textbf{M}_{\rm wedge}  \textbf{R}(\theta)\textbf{B}_{\rm trans} \textbf{S}.
\label{ConeBS}
\end{equation}
To recover \textbf{S} from an experimental system, the procedure outlined in equations~\ref{startAnalysis}-\ref{endAnalysis} is followed, replacing \autoref{Cone} with \autoref{ConeBS}, where the fundamental difference to the ideal case is the appearance of a $\cos2\theta$ term in the truncated Fourier series (see Supplement 1). The linear system of equations then becomes over-determined as we have five equations in only four unknowns. We therefore cannot use the direct matrix inverse to solve for \textbf{S} but instead use the matrix pseudo-inverse.

\section{Experimental Method}
\label{section4}

\begin{figure}[h]
\centering
\includegraphics[width=8.4cm]{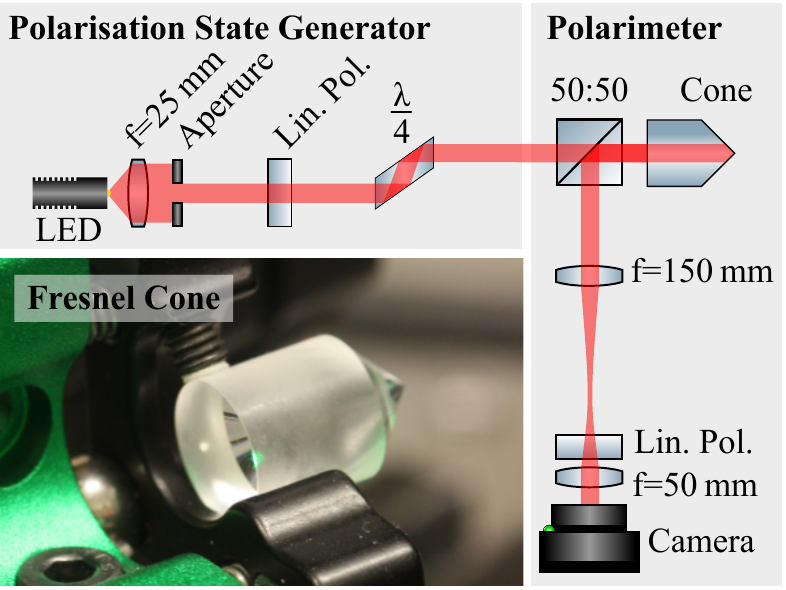}
\caption{Experimental setup. Polarisation states are generated on the left before entering the Fresnel cone based polarimeter on the right. Inset: Photograph of Fresnel cone with 10~mm diameter. }
\label{fig:setup}
\end{figure}

In a proof of principle experiment shown in~\autoref{fig:setup}, we demonstrate the operation of our Fresnel cone polarimeter for broadband light.  The system consists of a polarisation state generator (PSG), producing the initial polarisation states, and the actual polarimeter. The light source is a white LED, collimated by a lens (f=25.4 mm) and apertured to control the spatial coherence. The PSG consists of a linear polariser (Thorlabs, LPVISE100-A) and an optional $\lambda$/4 retardance Fresnel rhomb (Thorlabs, FR600QM) to generate broadband initial polarisation states. Half of the light then transmits through a non-polarising beam-splitter (Thorlabs, BS013) and reflects from a Fresnel cone (Edmund Optics 45-939, with aluminium coating removed). After back-reflection from the cone, half of the remaining light reflects from the non-polarising beam-splitter into the measurement arm, which consists of a horizontally aligned linear polariser (Thorlabs, LPVISE100-A) and lenses which re-image the plane of the cone tip onto a camera (Thorlabs, DCC1645C). The camera records the full profile of the reflected light from which we can extract $I(\theta)$. We note, however, that carefully placed individual detectors could be used instead. Resolving the zeroth, second and fourth azimuthal frequency components requires at least 5 detectors equally spaced in $\theta$, and distributed around half of the output beam. Using a camera allows us to obtain the intensity for quasi-continuous $\theta$ values (we use 499), and reduce the effect of noise in the system by averaging over the radial parameter.  In addition, we can extract the intensity at different colour channels, thus monitoring the operation across the visual spectrum.

\begin{figure}[h]
\centering
\includegraphics[width=8.4cm]{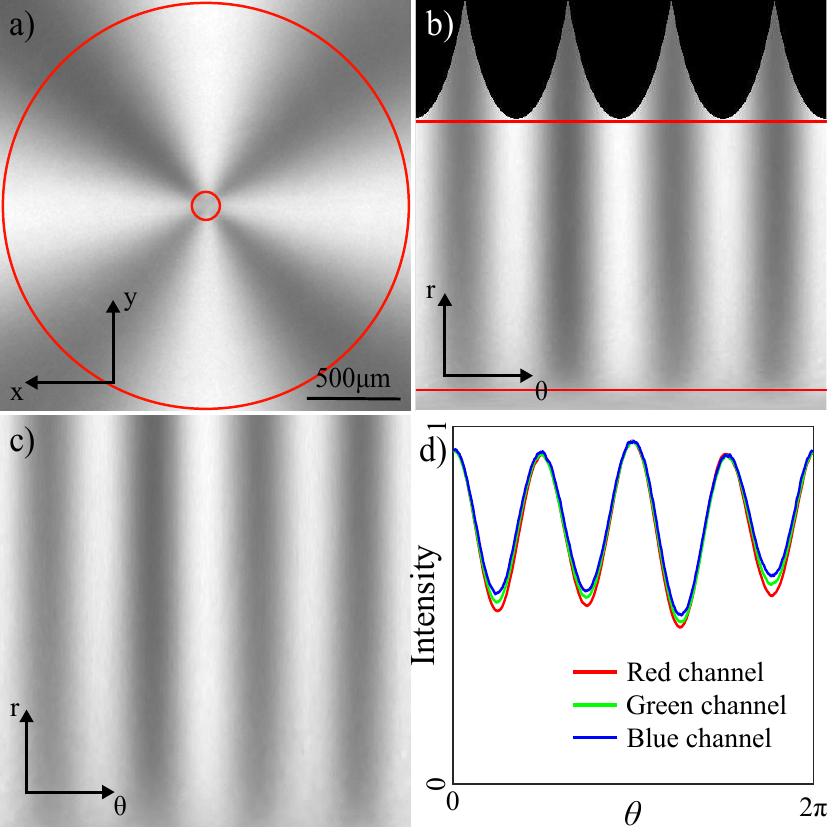}
\caption{Sample data for horizontally polarised initial light. a) Recorded intensity pattern with region of interest identified between the two red lines, ignoring the noisy area at the cone-tip. b) ``unwrapped'' intensity pattern from $x$-$y$ to $\theta$-$r$ coordinate frame and c) region of interest in $\theta$-$r$. d) Averaged 1D intensity profile shown for red, green and blue colour channels. a)-c) are intensities taken from the red colour channel.}
\label{fig:patterns}
\end{figure}

Light generated in different polarisation states produces different intensity patterns measured on the camera (an example for horizontal initial polarisation is shown in~\autoref{fig:patterns}a)). We calibrate the intensity of our images by recording a background image with the LED off, and two normalisation images with the PSG set to produce horizontal and vertical polarisation respectively. The intensity patterns used for our Fourier analysis are background corrected and normalised to the sum of the two normalisation images. We unwrap the calibrated images from a Cartesian ($x$-$y$) to a polar ($\theta$-$r$) coordinate frame, relative to the cone centre. The noisy data at the centre and edges of the intensity pattern is removed by selecting a region of interest, delimited by red lines in \autoref{fig:patterns}a) and \autoref{fig:patterns}b), leaving the cleaned data shown in \autoref{fig:patterns}c). 

The resulting averaged 1D intensity profile (see \autoref{fig:patterns}d)) is $I(\theta)$, the first row of~\autoref{ConeBS},  and from this we obtain the Fourier coefficients $\textbf{C}$ using a fast Fourier transform (FFT) as discussed in~\autoref{PolTheory}. Calculation of $\textbf{M}$ requires determination of $\textbf{B}_{\rm refl}$ and $\textbf{B}_{\rm trans}$, which we achieve following the method outlined in~\cite{Azzam1987,Fujiwara2007}. Systematic errors in these measurements can lead to non-physical Mueller matrices, which can be avoided by converting $\textbf{B}_{\rm refl}$ and $\textbf{B}_{\rm trans}$ into the form shown in Equation (5) of~\cite{Hauge1978} (see Supplement 1). This conversion assumes that there is no loss of polarisation in the beam-splitter and that the Mueller matrix can be parameterised by 3 angles, namely a phase-shift induced between s and p polarisation components, a rotation of the linear polarisation components, and an orientation of the beam-splitter from horizontal. We achieve this normalisation by numerical minimisation of the difference between our measured and normalised matrix while varying all 3 angles. The pseudo-inverse of $\textbf{M}$ is calculated using the singular value decomposition (SVD) method and we finally recover $\textbf{S}$ using \autoref{endAnalysis}.

The imperfections of the optical elements, described by the Mueller matrices $\textbf{B}_{\rm refl}$ and $\textbf{B}_{\rm trans}$, are in general frequency dependent. We identify these matrices for the red, green and blue colour channels of our camera independently. These are applied to the individual intensity patterns for the three colour channels, obtaining $\textbf{S}$ for red, green and blue frequency ranges. Our final white light polarisation state is taken to be their average. 

\section{Results and Discussion}
\label{ExpMethod}

\begin{figure}[h]
\centering
\includegraphics[width=8.4cm]{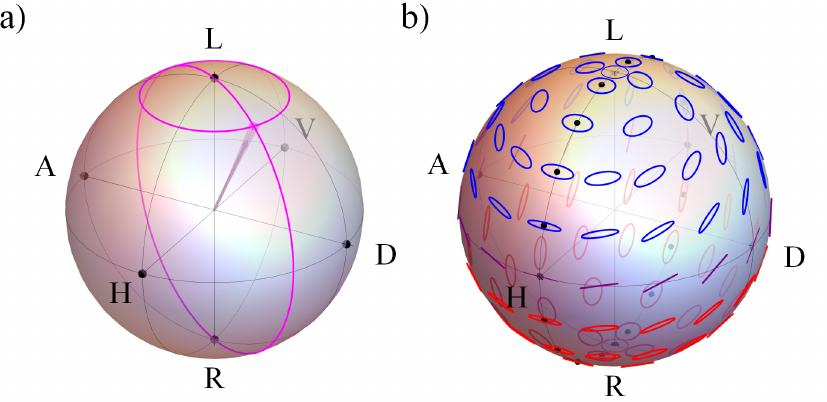}
\caption{Poincar\'e sphere diagrams showing a) example of the Fresnel cone polarimeter performance, where angular accuracy is represented by the solid cone (which in this example is set to $3.6^\circ$) and DOP is represented by the length of the vector within the Poincar\'e sphere (in this example set to 0.85), and b) dotted line where the dots represent which elliptical initial states are produced by the PSG in the elliptical case.}
\label{poincaresphere}
\end{figure}

To assess our polarimeter quantitatively, we consider its performance for measurement of the polarised and unpolarised components of the light separately. The polarised component is assessed by measuring the so-called angular accuracy, which can be thought of as the angle between initial and measured polarisation vectors on the Poincar\'e sphere as demonstrated in \autoref{poincaresphere}a). Angular accuracy is defined as 
\begin{equation}
\alpha=\cos^{-1} \left(\frac{\textbf{S}_{123}\cdot\textbf{S}'_{123}}{|\textbf{S}_{123}||\textbf{S}'_{123}|} \right),
\end{equation} 
where $\textbf{S}_{123}=[s_1,s_2,s_3]^T$ and $\textbf{S}'_{123}=[s'_1,s'_2,s'_3]^T$. The unpolarised component of the state can be quantified using the degree of polarisation (DOP) accuracy, where the DOP is defined as
\begin{equation}
\rm{DOP}=\sqrt{{s'_1}^2+{s'_2}^2+{s'_3}^2},
\end{equation}
and its accuracy is given by the magnitude of the difference between the initial polarisation state DOP and measured DOP. We assess the angular accuracy and DOP accuracy for a range of elliptical and linear initial states. The linear polarisation states are generated by rotating a linear polariser in $5^\circ$ increments over a range of $180^\circ$. Elliptical polarisation states are generated in the same way, with an additional quarter-wave Fresnel rhomb (horizontal fast-axis) after the linear polariser (see \autoref{poincaresphere}b) for a Poincar\'e representation of these polarisation states).

\begin{figure}[h]
\centering
\includegraphics[width=8.4cm]{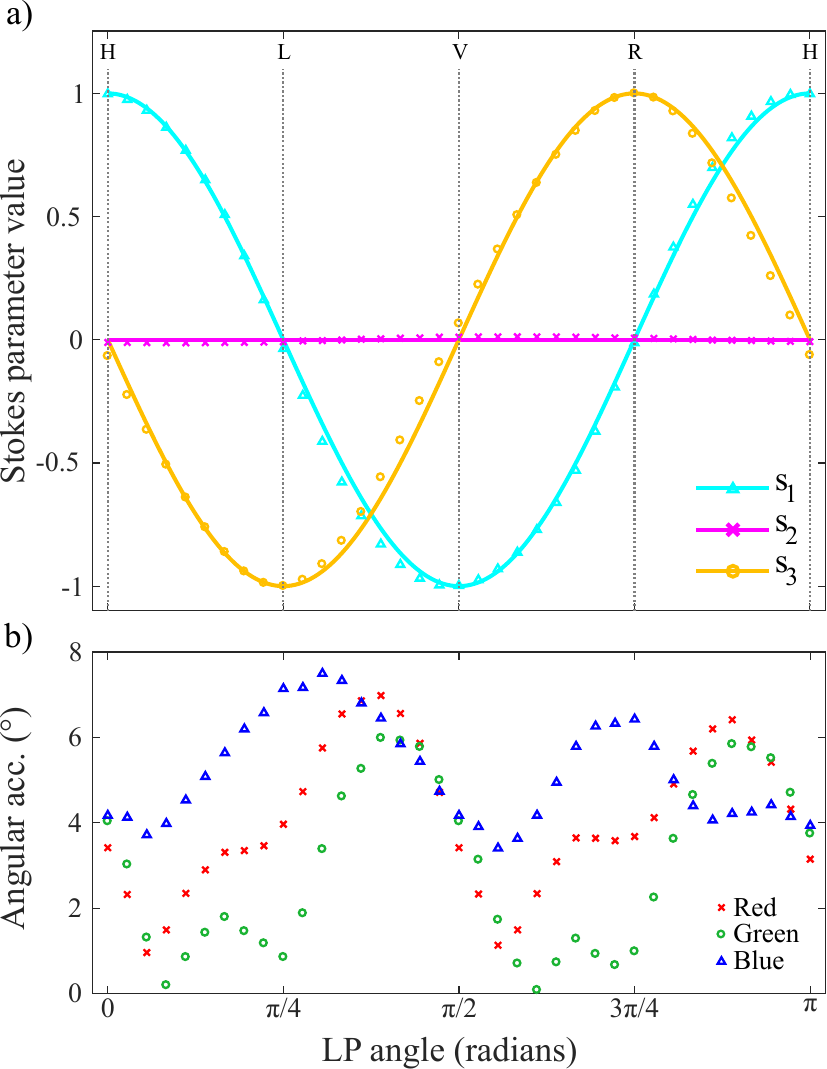}
\caption{Polarimeter performance for elliptical polarisation states, where a) shows ideal Stokes parameters (solid lines) with experimentally measured Stokes parameters (data points) for the initial polarisation states, and b) performance throughout the visible spectrum shown for the red, green and blue colour frequency bands.}
\label{doubleresults}
\end{figure}

To visualise the angular accuracy only, we set $s'_0=\rm{DOP}=1$. The remaining Stokes parameters for a range of elliptic input states are shown in~\autoref{doubleresults}a), finding excellent agreement between the initial (solid lines) and measured (data points) Stokes parameters. Similar results are obtained for the linear polarisation initial states (not shown). The overall performance can be gauged by taking the average angular accuracy for all elliptical (linear) initial states, which we find to be $2.9^\circ$ ($3.6^\circ$). 

In addition to the overall angular accuracy of the polarimeter we also investigate its frequency dependence, which is obtained by performing the data analysis on each colour channel of the camera individually. \autoref{doubleresults}b) shows the angular accuracy in the red, green and blue colour bands. We see that the polarimeter has similar performance across the three frequency bands for elliptical (linear) initial states, with average angular accuracies of $4.1^\circ$ ($2.1^\circ$), $3.0^\circ$ ($2.7^\circ$) and $5.2^\circ$ ($6.4^\circ$) in the red, green and blue colour bands respectively. These results confirm that our polarimeter performs well across the full visible spectrum.

%, minimum error of $0.03$($0.013$) and maximum error of $0.26$($0.08$).
%and is more important when measuring fully polarised states, as is often the case in applications. To visualise the angular accuracy we set $s'_0$ and DOP to 1 and show example data in \autoref{doubleresults}a), where we see an excellent agreement between initial and measured Stokes parameters.

%The above results are valid for all polarisation states including partial or unpolarised light. We next restrict ourselves to measurement of fully polarised states by setting  $s'_0$ and the DOP to 1, as this is often the case in application. \autoref{doubleresults}b) shows results when restricting to fully polarised states, and an excellent agreement between input and measured $s'_1$, $s'_2$ and $s'_3$ values is observed. We calculate angular accuracies of  for elliptical and linear input states, respectively. In both cases our polarimeter identifies the set polarisation state very well.

%To quantify these results, we first convert our Stokes vectors $(\textbf{S},\textbf{S'})$ into Jones vectors ($[E_{\rm x},E_{\rm y}]^T, [E'_{\rm x},E'_{\rm y}]^T$) and then calculate the fidelity $F$ using
%%
%\begin{equation}
%F=(E_{\rm x}E'^{*}_{\rm x}+E_{\rm y}E'^{*}_{\rm y})^2.
%\end{equation}
%%
%
%A fidelity of 1 is a perfect polarisation state match and 0 corresponds to orthogonal states. The average fidelity for all measurements of the generated linear polarisation states and elliptical polarisation states is $2.66$, with linear polarisations and elliptical polarisations having [1-LOG] fidelities of $2.58$ and $2.74$ respectively.

In addition to fully polarised states, our polarimeter also allows measurement of the full Stokes vector. We measure the DOP accuracy for the same elliptical (linear) initial polarisation states as before and calculate the average error to be $0.12$ ($0.08$) across all initial polarisation states and colour channels. Individual colour channel performance is 0.11 (0.10), 0.13 (0.07) and 0.11 (0.08) for the red, green and blue colour channels respectively, showing good performance across the visible spectrum.

The formal equivalence between our polarimeter and the canonical rotating quarter-wave plate polarimeter, enables us to apply the findings of previous optimisation studies. The number of measurements used for Stokes vector determination has been discussed in~\cite{Tyo2002a}, where it was shown that performance is increased when using more than the minimum required number of $\theta$ measurements. For temporally modulated polarimeters this requires a longer acquisition time, however here we record measurements for many $\theta$ angles simultaneously. In agreement with these studies we find best performance when we use as many $\theta$ values as possible (499). The optimum number of $r$ values however changes between anguar accuracy and DOP accuracy measurements. The best angular accuracy results are obtained by averaging all $r$ values from 25 pixels to 350 pixels, avoiding the noisy central region. Optimal DOP accuracy results are instead found when averaging over only 10 $r$ values at the edge of the image. We attribute this difference in optimal pixel range to the background noise level. At small $r$ there are fewer unique $\theta$ values due to pixellation, reducing contrast and increaseing backgroud noise. This effects the DOP measurements through its reliance on $a_0$, while leaving angular accuracy measurements relatively unafected.
% The main difference between the angular accuracy and DOP measurements is that DOP depends on $s_0$. Equation 5 shows that $s_0$ depends on $a_0$ - the zero frequency component or background light. $s_0$ and similarly DOP measurements are therefore more accurate when the contrast is maximised, which for our images occurs at the edge of the cone. The angular accuracy however is less sensitive to contrast and improves by averaging as many pixels as possible to reduce noise.

%If however we wished to minimise the number of measurements taken, the camera could be replaced by 5 individual photodiodes placed at locations of regularly spaced $theta$ values. Removing the requirement of using a camera would allow very fast acquisition rates, only limited by detector speed.

Performance can also be improved by adjusting the retardance, as discussed in~\cite{Sabatke2000b}, with an optimal retardance of $132^\circ$. The glass used in our Fresnel cone (BK7) has an average refractive index of $\sim1.52$ for white light, resulting in a retardance of approximately $79^\circ$. By engineering the Fresnel cone with refractive index of 1.86 (LaSF9 glass), a retardance of $132^\circ$ can be achieved, which is predicted to reduce the effect of measurement noise on the polarisation results. Compensation for errors in retardance value and azimuthal misalignment error of the linear polariser have also been discussed in~\cite{Flueraru2008a} and could be applied to the Fresnel cone polarimeter.

A potential advantage of our Fresnel cone polarimeter compared to conventional polarimeters is the increased acquisition rate. In our camera-based system, the data acquisition rate is set by the frame rate, allowing Stokes measurements in the kHz range using a suitable camera. Using individual detectors, as outlined in~\autoref{ExpMethod}, this could be increased into the MHz or GHz range. We can make an estimate of the performance of a setup with individual detectors by taking a subset of our camera data. Initial tests taking only 9 equally azimuthally spaced camera pixels (at $r=300$) around the cone shows an average angular accuracy of $5.4^\circ$ ($5.0^\circ$) and average DOP error of $0.12$ ($0.09$) for elliptical (linear) input polarisation states, confirming that the system still performs well even with few measurements.

\section{Conclusions}

We have demonstrated a full Stokes polarimeter based on the back-reflection from a Fresnel cone, and show that it is the spatial analogue to the ubiquitous rotating quarter-wave plate technique.  We characterise the performance of our polarimeter by measuring the angular accuracy, experimentally demonstrating this to have an average of $2.9^\circ$ and $3.6^\circ$, with average DOP errors of $0.12$ and $0.08$ for elliptical and linear polarisation states respectively. We have also shown that the polarimeter performs well across the visible spectrum, by measuring the angular accuracy and DOP error in the red, green and blue colour bands respectively. This proof of principle experiment uses off-the-shelf polarisation components, and specialised optics could improve performance while reducing size. Previous research on polarimeter optimisation can also be applied to our polarimeter, improving performance through optimisation of refractive index and measurement strategy.

Not only does our device record a full Stokes vector in a single shot for broadband light, but Fresnel cone polarimeters are robust, stable and low-cost ($\sim$£500). This new capability could find many uses in a wide range of applications, for example a multi-domain polychromatic spectro-polarimeter~\cite{Snik2014a,Snik2015a}, which provides Stokes measurements as a function of wavelength. While our polarimeter is not compatible with direct imaging, scanning techniques such as confocal or 2-photon microscopy could be equipped with our Fresnel cone polarimeter, adding polarisation sensitivity to these techniques.

\section*{Funding Information}

This work has been supported by EPSRC Quantum Technology Program grant number EP/M01326X/1. R. D. Hawley's work was supported by the EPSRC CDT in Intelligent Sensing and Measurement, Grant Number EP/L016753/1.

\section*{Acknowledgments}

We would like to thank Gergely Ferenczi and Jonathan Taylor for useful and interesting discussions.

\section*{Supplemental Documents}

See Supplement 1 for supporting content. A dataset containing the raw data and analysis code can be found at: http://dx.doi.org/10.5525/gla.researchdata.650.

%\bibliography{Polarimeter-Paper}

\end{document}